\documentclass{aa} 

\usepackage[varg]{txfonts}
\usepackage{graphicx}

\usepackage{multirow}

\newcommand{\degree}{\ensuremath{^\circ}}

\begin{document}

\title{ELISa: A new tool for fast modelling of eclipsing binaries}

\author{Michal \v{C}okina
   \and Miroslav Fedurco
   \and \v{S}tefan Parimucha
}
\institute{Department of Theoretical Physics and Astrophysics, Institute of Physics, Faculty of Science, University of Pavol Jozef Šaf\'arik, Park Angelinum 9, 041 54 Košice, Slovakia, \email{mikecokina@gmail.com}}

\date{Received \#Day Month \#Year / Accepted \#Day Month \#Year}

\abstract
{We present a new, fast, and easy to use tool for modelling light and radial velocity curves of close eclipsing binaries with built-in methods for solving an inverse problem.}
{The main goal of ELISa (Eclipsing binary Learning and Interactive System) is to provide an acceptable compromise between computational speed and precision during the fitting of light curves and radial velocities of eclipsing binaries. The package is entirely written in the Python programming language in a modular fashion, making it easy to install, modify, and run on various operating systems.}
{ELISa implements Roche geometry and the triangulation process to model a surface of the eclipsing binary components, where the surface parameters of each surface element are treated separately. Surface symmetries and approximations based on the similarity between surface geometries were used to reduce the runtime during light curve calculation significantly. ELISa implements the least square trust region reflective algorithm and Markov-chain Monte Carlo optimisation methods to provide the built-in capability to determine parameters of the binary system from photometric observations and radial velocities.}
{The precision and speed of the light curve generator were evaluated using various benchmarks. We conclude that ELISa maintains an acceptable level of accuracy to analyse data from ground-based and space-based observations, and it provides a significant reduction in computational time compared to the current widely used tools for modelling eclipsing binaries.}
{}

\maketitle

\section{Introduction}
\label{sec:intro}
The development in observational techniques and instruments in recent years has led to a vast amount of available data gathered from ground-based observations, especially from automated ground surveys such as SuperWASP \citep{Pollaco06}, PanSTARSS \citep{panstarrs2016}, ASASS \citep{asass2005}, OGLE \citep{ogle2015}, among others. In contrast, a rapid increase in the amount of observational data can be expected with prepared surveys such as Rubin Observatory (LSST) \citep{rubin2019}. Space-based observations from observatories such as  Kepler \citep{Borucki16}, TESS \citep{Ricker15}, or GAIA \citep{gaia16} have significantly contributed to a huge library of photometric observations of an unprecedented quality and duration. The vast majority of gathered data consists of the light curves (LCs) of celestial objects, including the LCs of eclipsing binaries (EBs). Photometric data represent one of the primary sources of knowledge we have about objects, such as eclipsing binaries. Continuous improvement in the accuracy of observational data needs to be supplemented with the development of suitable processing and modelling capabilities that enable us to infer parameters of the object from observations and fully utilise the precision of the observational data. The amount of the LCs produced by such instruments means that the researcher can analyse only a tiny fraction of observations with the standard customised object-to-object approach. Therefore, to significantly increase the number of analysed EBs, efficient and primarily autonomous solutions to the inverse problem need to be developed.

There are numerous software packages available dedicated to the LC modelling of EBs. The majority of the modelling tools such as PHOEBE v1 \citep{prsa01} or ELC \citep{Orosz00} are based on the Wilson-Devinney (WD) code \citep{WilsonDevinney1971}. In addition, other LC integrators are using a different approach, for example, using triaxial ellipsoids to model binary components as utilised in the package ELLC \citet{Maxted16}. The abovementioned packages, although fast, have their own limitations in terms of precision, which are discussed in section \ref{sec:purpose}. Probably the most advanced binary modelling tool available is the package PHOEBE v2 \citep{prsa02}, which encapsulates a huge variety of functionalities while achieving exceptional precision, however, at the price of relatively slower runtime speed compared to the previously mentioned packages.

\section{Main purpose of the ELISa package}
\label{sec:purpose}
ELISa (Eclipsing binary Learning and Interactive System) is a cross-platform software package dedicated to modelling EBs, including surface features such as spots and eventually pulsations. The package is written in the Python programming language, and it is freely available online\footnote{https://github.com/mikecokina/elisa}.

The presented piece of software is an easy to install pure Python package using a modular structure that enables rapid implementation of new features to the existing framework. The package has been tested and developed on Windows and Linux operating systems. It also utilises a modern approach to EB modelling, emphasising computational speed while maintaining a sufficient level of precision to process modern ground-based and space-based observations. The primary purpose of this package is to allow for one to infer parameters of the eclipsing binaries from LC and radial velocity (RV) observations utilising the least square trust region reflective (LSTRR) algorithm and the Markov-chain Monte Carlo (MCMC).

ELISa is specialised for tasks where a huge number of synthetic observations need to be evaluated in a reasonable amount of time for applications such as machine learning or solving an inverse problem, that is to say inferring parameters of EB from observational data. The package was developed using a modern high-level programming language, Python. Therefore, users can also write scripts in pure Python, which enables one to utilise a vast amount of software tools available in Python, making this package suitable for applications where a significant level of automation is needed. The package functions are written according to standard coding practices used in Python, which should ensure an easy learning process for users with a rudimentary knowledge of the Python programming language. The software package is also accompanied by a comprehensive set of Jupyter notebook tutorials that guides the user through the basics of using the ELISa package.

As was mentioned in the introduction, numerous software packages dedicated to modelling binary stars have been published, achieving a decent runtime speed. However, one of the shortcomings of the WD-derived code (e.g. WD, PHOEBE v1, and ELC) is the use of trapezoidal discretisation. In this case, the stellar surface is represented as a discrete set of points, where physical properties are calculated. As a result, WD can produce inaccuracies in the calculated integrated flux, especially during eclipses. There is also the issue with the assignment of corresponding areas to each surface point. In deformed surfaces,  surface point density does not stay constant across the surface, which can pose issues during a light curve integration and treatment of effects, including deformations of the stellar surface, such as non-radial pulsations. Modelling approaches departing from the use of Roche geometry, such as in the package ELLC, are not adequate in the case of the close binaries where the approximation of the stellar surface by triaxial ellipsoid necessarily breaks down. ELISa addresses the abovementioned issues and still reduces the computational time compared to the more sophisticated codes such as PHOEBE (see section \ref{sec:precision}).

The precision and run-time speed are software requirements that are usually mutually exclusive. ELISa using suitable and widely used computational tools (such as NumPy and SciPy) as well as the unique features of EB modelling, which are described in detail in this paper, can provide an acceptable compromise between those two requirements. ELISa offers an opportunity to use built-in fitting methods to solve an inverse problem. Vast amounts of observational data available means that the package should require minimal user input to perform a rudimentary analysis of the EB. Users can use a quick LC generator to generate extensive learning data sets for machine learning applications in EB research.

\section{Physics of ELISa}
\label{sec:physics}
Basic principles behind the determination of the surface geometry, temperature distribution, and radiation properties of the surface follow general principles used in the WD code. Equilibrium surfaces of the components are generated as equipotential surfaces utilising Roche geometry. This approach is suitable for binary systems with an eccentric orbit, asynchronous rotation, and the rotation axis of the components perpendicular to the orbital plane \citep{wilson79}. 

Compared to the original WD code, there are numerous improvements, primarily to improve the precision of calculated light curves, especially in close eclipsing binaries. The modifications can be summarised in the categories described in the Sections

\section{Surface discretisation}
\label{sec:surf_points}
The shape of the binary components is defined by solving an equation for binary potential generalised to account for eccentric and asynchronous orbits (see equation 1 in \citet{wilson79}). The potential function is used to produce a discrete set of surface points, creating a mesh that defines surface geometry. This set of points should cover the whole stellar surface homogeneously and with sufficient fidelity to not produce any unwanted artefacts during the curve modelling. The density of surface points in ELISa is governed by the discretisation factor, representing the mean angular distance between the surface points.

Trapezoidal discretisation used in WD code is not particularly suitable for close binaries due to a substantial deformation of the component. The mentioned method is especially unsuitable in the case of surfaces of 'overcontact' binaries \citep{Wilson01} due to the presence of the 'neck' connecting the components. Detached and overcontact binaries use different approaches for surface discretisation due to significant differences between their geometries. Both discretisation methods are described in the following sections \ref{sec:detached_points} and \ref{sec:over_contact_points}.

\begin{figure}
	\centering
		\includegraphics[width=0.49\textwidth]{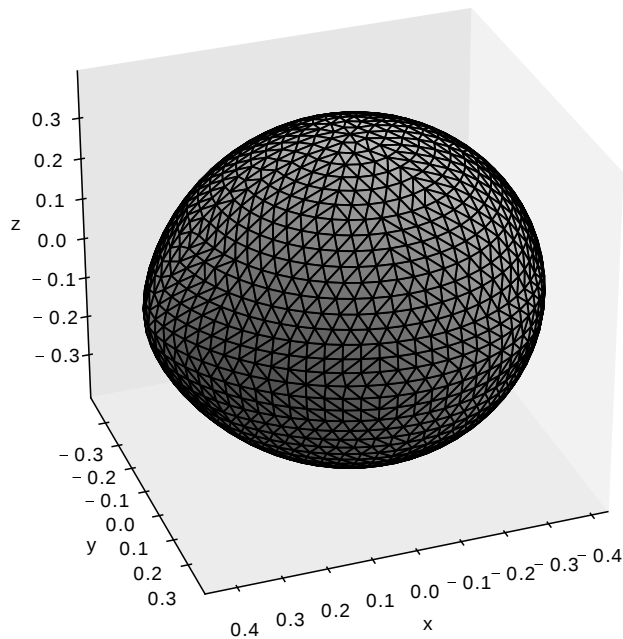}
		\caption{Surface of the detached system generated using discretisation factor 5.}
		\label{fig:surface_discretization_detached}
\end{figure}

\subsection{Detached systems}
\label{sec:detached_points}
Surfaces of detached binaries use an improved form of a trapezoidal discretisation where corrections to original trapezoidal discretisation are applied to angular coordinates $\varphi_{t}$ and $\vartheta_{t}$:
\begin{equation}
\label{eq:detached_trapezoidal_corr}
\begin{split}
 \vartheta &= \theta_t + \arctan{\frac{(\overline{r}_{eq} - r_{polar})\tan{\vartheta_t}}{r_{polar} + \overline{r}_{eq}\tan^2{\vartheta_t}}}, \\
 \varphi &= \varphi_{t} + \arctan{\frac{(r_{side} - r_{point})\sin{\vartheta}\tan{\varphi_t}}{r_{side} + r_{point}\tan^2{\varphi_t}}};~~~~\varphi_t \in \left( 0, \frac{\pi}{2} \right), \\
 \varphi &= \varphi_{t} + \arctan{\frac{(r_{side} - r_{back})\sin{\vartheta}\tan{\varphi_t}}{r_{side} + r_{back}\tan^2{\varphi_t}}};~~~~\varphi_t \in \left( \frac{\pi}{2}, \pi \right),
\end{split}
\end{equation}
where $\overline{r}_{eq}$ is an 'average' equatorial radius estimated as the mean of point, side, and back radii ($r_{point}$, $r_{side}$, and $r_{back}$). The example of the discretisation for a detached binary surface is Figure \ref{fig:surface_discretization_detached}.

\begin{figure}
	\centering
		\includegraphics[width=0.49\textwidth]{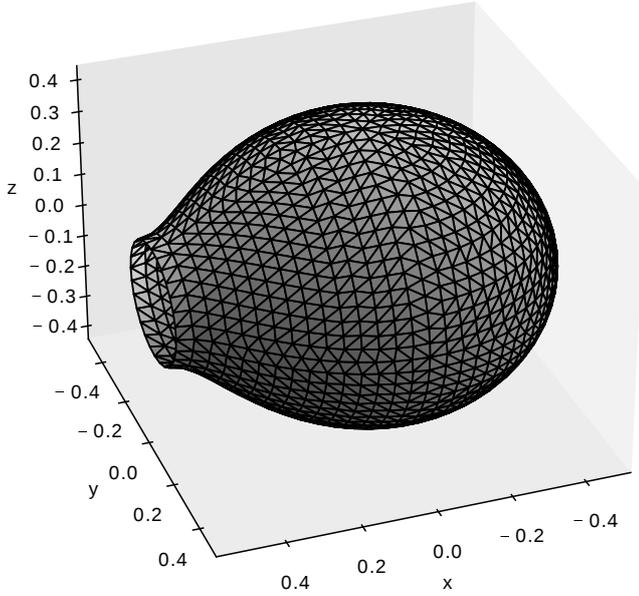}
		\caption{Surface of the overcontact system generated using discretisation factor 5.}
		\label{fig:surface_discretization_over_contact}
\end{figure}

\subsection{Overcontact systems}
\label{sec:over_contact_points}
The previously described method is not suitable for the discretisation of overcontact surfaces. The presence of the neck causes the corrected trapezoidal method to break down. Therefore, due to the difference in geometry, the overcontact surface is split into the near-side and far-side parts. The angular coordinates for the far-side points are generated in the same manner as in the detached surface. On the other side, a different discretisation method was used for the near-side points forming the neck. The discretisation can use cylindrical symmetry along an $x$-axis with much better results. The surface points are arranged in concentric rings around the $x$-axis, spaced equidistantly along the curved stellar surface on the $\varphi = 0$ meridian. The cylindrical angular coordinate $\varphi_{c, linear}$ measured from the direction of the Cartesian $z$ coordinate, where $z_c \equiv x$, is then generated equidistantly along generated concentric rings. Finally, the $\varphi_{c, linear}$ is then corrected for tidal deformation using the following correction:
\begin{equation}
 \label{eq:corr_contact_binary}
 \varphi_c = \varphi_{c,linear} + \arctan{\frac{(\delta_r - 1)\tan{\vartheta}}{1 + \delta_r\tan^2{\vartheta}}}; ~~~~ \delta_r = \frac{r_{side}}{r_{polar}}.
\end{equation}
The resulting surface discretisation for the overcontact surface is displayed in Figure \ref{fig:surface_discretization_over_contact}.

\subsection{Calculating radial coordinates}
The desired $N$ stellar surface points can be obtained by solving the binary potential equation for the radii {\boldmath$r$} corresponding to the given set of angular coordinates {\boldmath$\theta$}, {\boldmath$\phi$} generated with roughly the same angular separation in sections \ref{sec:detached_points} and \ref{sec:over_contact_points}. Near-side points of the overcontact system components were generated in cylindrical symmetry instead of spherical symmetry to ensure proper surface coverage of the points on the neck. Since the equation for the binary potential has to be solved for {\boldmath$r$} implicitly, the surface point generator can use a suitable numeric solver such as the Newton solver, where substantial saving in computational time can be achieved by using a vectorised approach. The coordinates of the surface points are solved as an N-dimensional vector {\boldmath$r$}, {\boldmath$\theta$}, {\boldmath$\phi$}.  Thus, the implicit solver has to be called just once instead of N times. The number of surface points N depends on the discretisation factor defined by the user, and it represents the mean angular size of the surface elements in degrees. The lower value of the discretisation factor results in a higher number of surface elements.

\subsection{Utilisation of surface symmetry}
\begin{figure}
	\centering
		\includegraphics[width=0.49\textwidth]{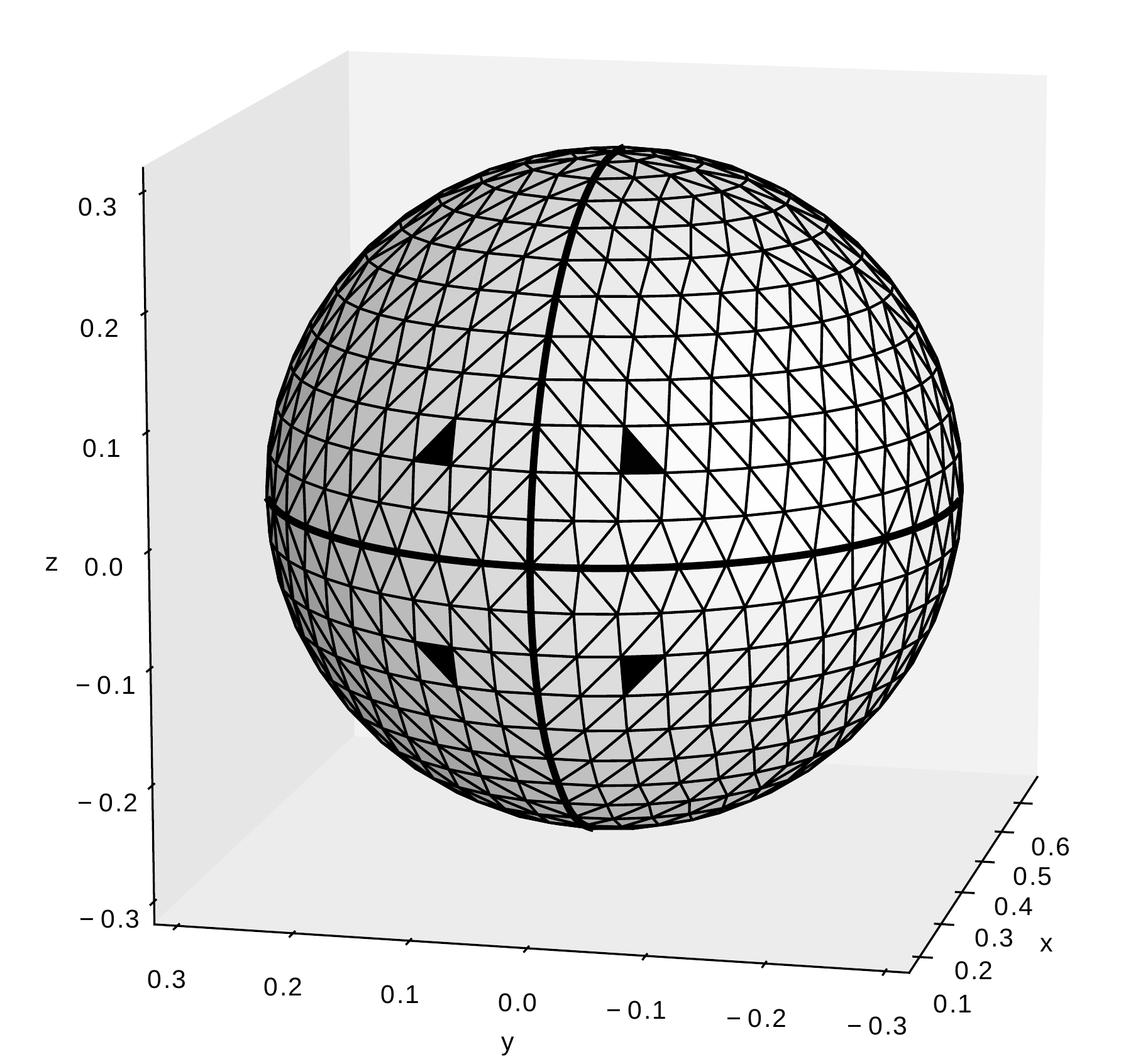}
		\caption{Visualisation of the surface symmetry of the EB component. Bold curves divide the surface into four symmetrical parts and mark an intersection of symmetry planes with the component's surface. Black triangles identify mutually symmetrical surface elements, which can share the same surface properties.}
		\label{fig:symmetry}
\end{figure}

Further improvement to the computational speed can be achieved by utilising planar surface symmetries of the eclipsing binary surface where any equipotential surface can be divided into four symmetrical parts. Therefore, the surface points generator needs to calculate only roughly N/4 points. The rest of the points are symmetrical images obtained by planar reflection along $xy$ or $xz$ planes in the co-rotating frame of reference. Such a coordinate system is anchored to the centre of mass of the component in question, while $x$-axis points towards the centre of mass of the companion component (see Figure \ref{fig:symmetry} for details).

\section{Triangulation of surface points}
\label{sec:triangulation}
Surface points obtained in section \ref{sec:surf_points} were subsequently tessellated using triangular faces whose vertices are surface points obtained in Section \ref{sec:surf_points}. This approach ensures an increased level of precision compared to the WD code, and it is commonly used in computer graphics and packages dedicated to EB modelling such as PHOEBE v2 \citep{prsa02}. Since the component's surface geometry significantly depends on the system morphology, the triangulation process differs slightly for detached and overcontact systems.

\subsection{Detached binaries}
\label{sec:dtbt}
Surfaces of components in detached binaries are from geometrical perspective convex objects in three-dimensional space. Probably one of the most effective triangulation methods for convex objects is the Delaunay triangulation algorithm \cite[][]{delaunay01}. The Delaunay triangulation in two dimensions is based on the criterion of empty circumcircles of triangles creating a triangulated convex hull. The two-dimensional version of this algorithm can be adapted in three dimensions where the circumsphere for the given tetrahedron should not contain any additional points. The desired surface is then obtained as the convex hull of such a triangulation. Therefore, the method can be used directly to describe the surface of the detached binary component. 

\begin{figure}
        \centering
                \includegraphics[width=0.49\textwidth]{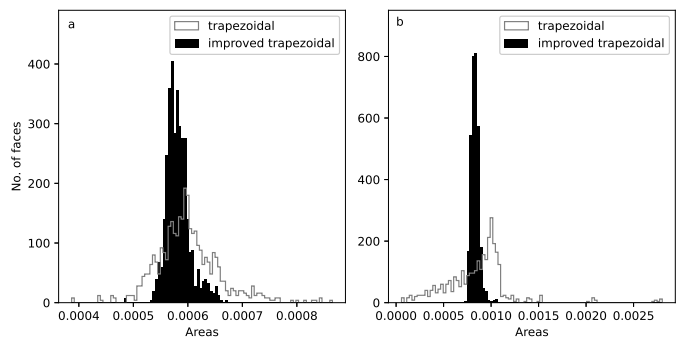}
                \caption{Distribution of the surface element areas for the detached system (panel a) and overcontact system (panel b). Each panel contains distributions for the original trapezoidal discretisation and improved trapezoidal discretisation method.}
                \label{fig:area_histograms}
\end{figure}

Maintaining the size of the surface elements as similar as possible is crucial to the efficient and precise description of the stellar surface. Similar-sized surface elements ensure the adequate treatment of the surface parameter gradients. Additionally, the process that rectifies the underestimation of the stellar surface discussed in Section \ref{sec:horizon_est} requires similar-sized surface elements to function as intended. With significant differences in surface element sizes, the level of surface underestimation would vary too much across the surface and thus unnecessarily complicate the solution to this issue. Therefore, the surface discretisation process used corrections to the trapezoidal discretisation listed in Equation \ref{eq:detached_trapezoidal_corr} to reduce the variance in the surface element sizes. The effect of the mentioned corrections on the distribution of surface element areas is examined in Figure \ref{fig:area_histograms}. We can conclude that the corrections listed in Equation \ref{eq:detached_trapezoidal_corr} managed to reduce the dispersion of the surface element area distribution from 3.5 to 1.5\% of the mean surface element area based on the half-width of the (16, 84) percentile interval. This reduction helped maintain the dispersion in the conservation of the flux benchmark in Figure \ref{fig:solar_const} within the targeted precision of the numerical model.

As in Section \ref{sec:surf_points}, the size of the triangulated surface without surface inhomogeneities (e.g. spots or pulsations) can be reduced to its symmetrical part since the convex nature of the remaining symmetrical section of the surface is maintained. Subsequently, triangulation can be simply mirrored to the rest of the surface. Figure \ref{fig:symmetry} demonstrates the use of surface symmetries where the example of mutually symmetrical faces are coloured black.
\subsection{Overcontact binaries}
The surface of overcontact binaries significantly differs from their detached binary system counterparts. Both components are physically connected with the so-called neck, which is not a convex object. Thus, the surface of either component cannot be found with the Delaunay triangulation. Several numerical methods can be used for triangulation of concave surfaces, such as the Poisson surface reconstruction \citep{Kazhdan06poissonsurface} or marching method described by \citet{Hartmann98}. These methods are implemented in various, primarily graphic software libraries. However, ELISa utilises the adapted form of Delaunay triangulation to deal with the overcontact binary surfaces to use its efficiency. 

The main idea of our adaptation of the Delaunay triangulation algorithm to over contact surfaces is a suitable transformation of the original surface points into a convex object, while maintaining relative positions of transformed surface points. Subsequently, the standard Delaunay triangulation can be used to triangulate the transformed points. Finally, the resulting simplices (i.e. indices of triangle vertices) can be applied to the original set of surface points.

\begin{figure}
        \centering
                \includegraphics[width=0.49\textwidth]{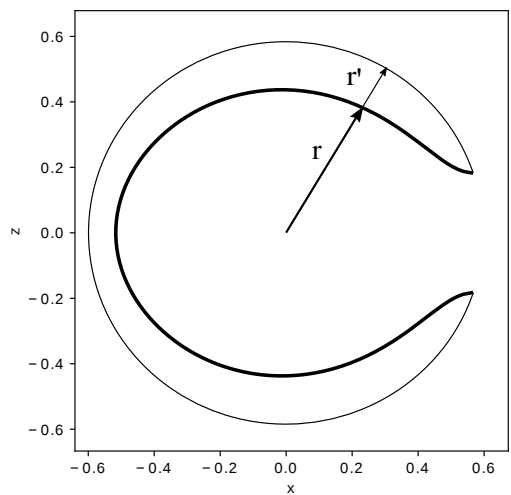}
                \caption{$xz$ cross-section of the transformation from over the contact component surface (thick inner curve) to convex shape suitable for Delaunay triangulation (thin outer curve). Transformations defined by Equation \ref{eq:transform_over_contact} are illustrated by position vector $\mathbf{r}$ and its transformed counterpart $\mathbf{r^\prime}$.}
                \label{fig:over_contact_transformation}
\end{figure}

In our transformation, surface points $\mathbf{r}$ are transformed into a sphere with a radius equal to the distance of the component's neck to the centre of mass of the star $x_{neck}$ using the following transformation:
\begin{equation}
\label{eq:transform_over_contact}
\mathbf{r^{\prime}} = x_{neck} \frac{\mathbf{r}}{|\mathbf{r}|}.
\end{equation}
The spherical transformation described in equation \ref{eq:transform_over_contact} is visualised in Figure \ref{fig:over_contact_transformation} and the resulting surface can be seen in Figure \ref{fig:surface_discretization_over_contact}.

Surface symmetry can also reduce the amount of transformed surface to triangulate as discussed in Section \ref{sec:dtbt}. The triangulation process tessellates the surface of the overcontact binary without any gaps or overlaps on the neck thanks to how the surface discretisation generated surface points (see Section \ref{sec:surf_points}). The transformation described above is not the only viable one. In general, any transformation converting an overcontact surface to a convex object should work.  

We have also investigated the distribution of the surface element areas in the overcontact surfaces. The resulting histogram is displayed in the Figure \ref{fig:area_histograms}b. The improved trapezoidal method drastically reduced the dispersion of surface element sizes compared to the original trapezoidal discretisation. The root cause of the large dispersion was the original trapezoidal discretisation method assigning disproportionately large areas to surface elements located on the neck. The use of cylindrical symmetry on the neck points combined with the calculation of the surface areas on triangulated surface elements substantially decreased the variation of the surface element sizes.
\section{Modelling of spots in ELISa}
\label{sec:spot_modelling}
ELISa treats stellar spots as inhomogeneities in surface temperature distribution. The temperature inhomogeneities can be caused by magnetic activity, such as in the case of variables of an RS CVn type \citep{Pribulla00, Pribulla01, Vanko07}. Additionally, variations in the local chemical composition of the stellar surface, such as in the case of chemically peculiar stars \citep{Strassmeier17, Paunzen18}, can lead to spot variability. In the case of magnetic field-driven spots, the disturbances in a local magnetic field of the star with a convective envelope is capable of generating local suppression of convection that transfer energy from deeper parts of the star towards the surface. Such places usually appear colder and thus darker than surrounding areas. Hotter regions may also appear on the surface due to activity in the photosphere of the star.

On the other hand, chemical spots are presumed to be caused by the radiative levitation of certain heavy elements in the case of slowly rotating chemically peculiar (CP) stars with radiative envelopes \citep{Michaud76}. The chemical spots are a probable cause of photometric variability due to the redistribution of flux in the affected parts of the stellar surface that would lead to rotational spot variability \citep{Hummerich18}. As a first approximation, the chemical spots are currently modelled inside the framework of ELISa as regular temperature spots as in the case of spots driven by magnetic field until the mechanism of CP stars variability is fully understood.  

Spots are, in some cases, necessary to include in the EB modelling procedure to adequately model the observations. One of the suitable examples is the O'Connell effect, where the presence of the spots in the EB model is necessary to explain asymmetries in LCs of certain close EBs. The package can also model spots with small angular radii without increasing the fidelity of the surface discretisation globally, which would otherwise significantly increase the computational time needed for the calculation of the synthetic observation.

\begin{figure}
 \includegraphics[width=1.0\linewidth]{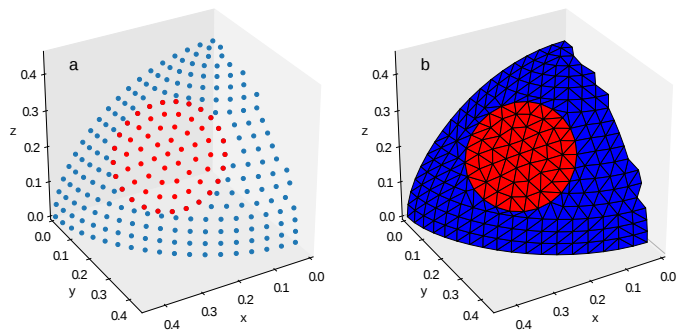}

 \caption{Surface mesh of a binary system component with a single spot (a) and the resulting triangulation (b). Circular symmetry of a discretisation enables one to maintain a well-defined edge between a spot and the rest of the surface.}
 \label{fig:spot_discretization}
\end{figure}

Spots in ELISa are modelled as circular regions on the stellar surface with a surface temperature defined by a temperature factor $T_{spot}/T_{star}$. Its angular radius determines the size of the spot, and the position of the spot centre is defined by the longitude measured from the join vector between components in a counter-clockwise direction and latitude measured from the north pole. Surface discretisation described in Section \ref{sec:surf_points} would, unfortunately, produce an irregular spot boundary. Therefore, surface points were generated for each spot in a manner demonstrated in Figure \ref{fig:spot_discretization}. Surface points of the spot were generated concentrically around the centre of the spot, keeping a clear circular boundary between the spot and the surrounding stellar surface. Subsequently, the underlying points from the original discretisation were discarded, and the resulting surface points, including spot points, were triangulated to form the resulting surface (see Figure \ref{fig:spot_discretization}). An iterative process was used for multiple overlapping spots that can be layered on the surface (and upon previously defined spots) to produce more complex temperature inhomogeneities. Surface points containing spots were generated on the initial point cloud well before the surface triangulation and calculation of the surface quantities. Therefore, a time penalty connected with the partial re-sampling of the surface points was kept minimal since the time necessary to generate surface mesh is insignificant compared to the total computational time required to produce a complete model of a binary.

Well defined spot boundaries are essential in order to perform modelling of small spots comparable in size to the used discretisation factor. Relatively small spots defined with surface elements from the original surface can lead to a sudden and significant change in the spot's total area. The issue can be caused by either a difference in the spot's location (asynchronous rotation) or by a different surface geometry since such a spot would cover only a handful of surface elements with an uneven boundary. Modelling small spots can also be solved using the original surface discretisation accompanied by a determination of partial coverage of the surface elements by the spot itself, similar to the algorithm for determining a partially covered surface element during an eclipse in Section \ref{sec:visibility}. However, such an approach creates a boundary region at the edge of the spot with the width being dependent on a discretisation factor.

The main disadvantage of the spot discretisation method discussed in this section is a significant variation in surface element sizes on the spot-star boundary (Figure \ref{fig:spot_discretization}b). Consequently, this feature locally deteriorates the homogeneity of the surface discretisation, which slightly deteriorates a surface's numerical precision. In conclusion, further research on implementing a spot discretisation on a stellar surface is needed to evaluate the benefits of different approaches. In ELISa, the authors prioritised the ability to produce a well defined star-spot boundary over the local homogeneity of the surface discretisation.

\section{LC integration techniques}
\label{sec:lc_integration}
Once the model of EB has been built, the radiant intensity of a binary system in the direction of the observer needs to be calculated by summing contributions of $N^\prime$ visible surface elements. The identification of visible surface elements consists of two steps. Initially, the surface elements facing towards the observer are identified using the angle between the outward-facing normal of the surface element and the line of sight vector. Subsequently, the visibility of the surface elements facing the observer during the eclipse is determined using the algorithm discussed in detail in Section \ref{sec:visibility}. In general, this process needs to be repeated for each orbital position in which an LC is to be calculated. However, different types of close EBs enable various optimisation levels of an LC integration technique that can provide a substantial reduction in computational time. Multiple approaches to the LC integration for circular and eccentric orbits are described in the following subsections.

\subsection{Circular orbits}
\label{sec:circular_lc_integration}
Properties of the close EBs with circular orbits can substantially accelerate the LC integration of such objects. In the simplest case, that is an EB without spots, the LC symmetry allows one to calculate only one-half of the required LC points while the rest can be mirrored around the primary or secondary eclipse. The next improvement can be used when surface features on the EB model stay constant in the co-rotating frame of reference, such as in the case of EB components without spots or the case of the EB model with synchronously rotating components with spots. In such instances, a single EB model can be used to calculate the LC points in each orbital position by simple rotation of the EB model. In a remaining case of asynchronously rotating EB components with spots, each spot needs to be re-incorporated into a clean surface for each orbital position and, thus, an overall reduction of the computational time is very much diminished in such case. 

\subsection{Eccentric orbits}
\label{sec:eccentric_lc_integration}
The standard method of the LC evaluation of the EB model with eccentric orbit suffers from the necessity to recalculate surface geometry due to a change in a component's distance during the orbital motion. Additionally, the corrections to the surface potentials of the components need to be derived by numerical means to conserve volumes of the components. These issues significantly slow down the evaluation of LCs compared to EBs with circular orbits discussed in Section \ref{sec:circular_lc_integration}. However, the following approximations applicable for EBs can significantly accelerate the calculation of an LC while maintaining a level of precision reached by the surface discretisation process (see Figure \ref{fig:lc_precision}).    
\begin{figure}
        \centering
                \includegraphics[width=0.99\columnwidth]{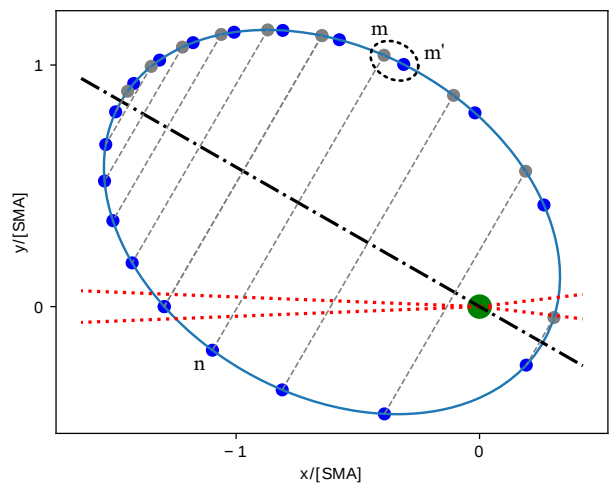}
                \caption{Visualisation of the orbital symmetry along the apsidal line (dash-dotted line) demonstrated a case of LC integration from a set of user-defined orbital positions (blue dots). Symmetrical orbital positions (grey dots) share binary models with symmetrical counterparts on the opposite ends of the grey dashed lines. Red dashed lines indicate positions of eclipse boundaries. The circled pair of orbital positions share the same surface geometry in the symmetrical counterparts approximation discussed in Section \ref{sec:second_approx}. The orbital positions are under-sampled for the sake of clarity.}       
                \label{fig:orbit_symmetry}
\end{figure}
\begin{figure*}
        \centering
                \includegraphics[width=0.9\textwidth]{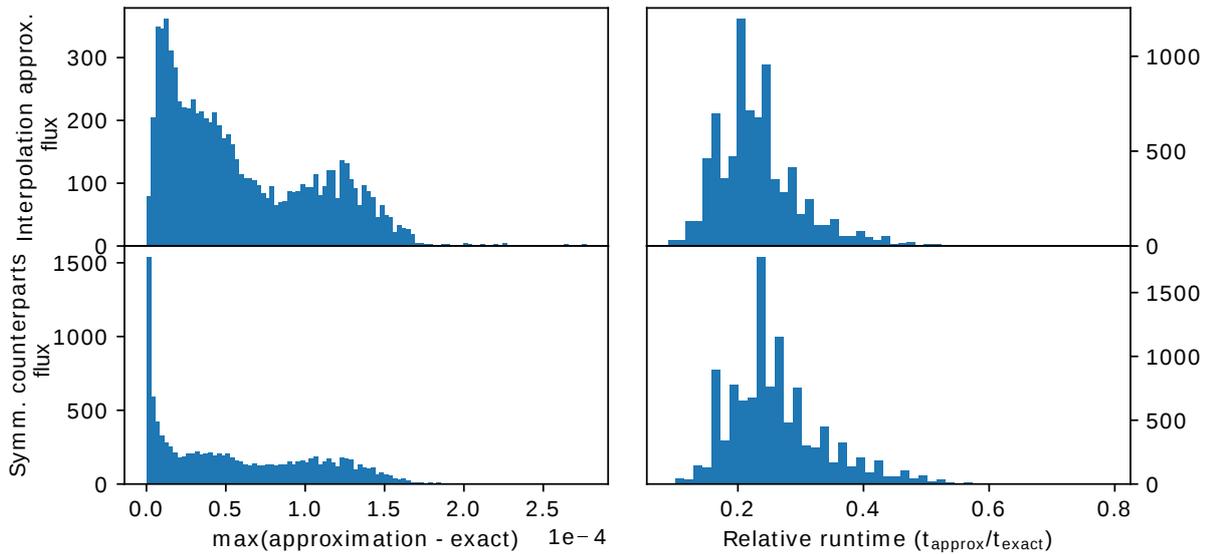}
                \caption{Histograms show precision (left column) and relative speed (right column) of the approximations used during the calculation of the LCs of EBs with eccentric orbits. The interpolation approximation (Section \ref{sec:first_approx}) in the top row and the symmetrical counterparts approximation (Section \ref{sec:second_approx}) located in the second row were tested separately. Both runs also utilised the similar neighbour approximation (Section \ref{sec:third_approx}). A benchmark consisting of 10\,000 randomly generated binary systems evaluated the precision and speed of both approximations. The process of generating the data set is discussed in detail in Section \ref{sec:approx_pecision_vs_speed}. Histograms demonstrate that the constraints put in place during the evaluation of approximations were able to constraint the maximum value of residuals below $2\times 10^{-4}$, while providing a significant improvement to the computational speed.
                }       
                \label{fig:ecc_lc_approx}
\end{figure*}
\subsubsection{Interpolation approximation}
\label{sec:first_approx}
The first approximation utilises the symmetry of an eccentric orbit along its apsidal line. In such a case, the surface geometries and thus the EB models are identical for the orbital positions $n$ and $m$ if their true anomalies $\nu_n$ and $\nu_m$ satisfy the following condition:
\begin{equation}
    \label{eq:ecc_orb_symmetry}
    \nu_m = 360\degree - \nu_n.
\end{equation}
The symmetry in this equation \ref{eq:ecc_orb_symmetry} enabled us to build a single EB model and we used it to calculate the LC points at the orbital positions $n$ and $m$ just by rotating the same EB model to their respective orbital positions. However, the main issue arises from the fact that the orbital positions of the desired LC points are generally not aligned symmetrically around the apsidal line as visualised in Figure \ref{fig:orbit_symmetry}. In the interpolation approximation, this issue is solved by calculating the EB models precisely on one side of the apsidal line. These EB models are then used to calculate the symmetrical LC points on the opposite side of the apsidal line (grey points in Figure \ref{fig:orbit_symmetry}). Finally, the LC points at the desired orbital positions are interpolated using the previously calculated symmetrical LC points. The shortcoming of this approximation is a requirement of sufficient density of the LC points to reduce the errors introduced by the interpolation process, mainly during eclipses. Therefore, the approximation is implemented only in cases where true anomalies of adjacent orbital positions do not exceed a pre-determined angular separation, which we found to be 0.08\, rad. The value is based on the evaluation of large sets of randomly drawn eclipsing binaries in a process similar to one described in \ref{sec:approx_pecision_vs_speed} using different values of the maximum allowed angular separation.

\subsubsection{Symmetrical counterparts approximation}
\label{sec:second_approx}
In this approximation, the orbital position counterpart and the closest user-defined orbital position are grouped to form orbital position couples (e.g. the pair of orbital positions circled in \ref{fig:orbit_symmetry}). Subsequently, the same model of a binary is assumed for both orbital positions in the couple. Compared to the approximation in Section \ref{sec:first_approx}, this approximation improves the precision during eclipses since the symmetrical LC point is calculated directly from the EB model at the desired orbital position instead of being interpolated. The approximation of symmetrical counterparts generally performs better for high eccentricity orbits ($e$>0.5) since it does not create artefacts at the edge of the eclipse as the interpolation approximation. The main issue of this approximation is that the paired orbital positions do not perfectly satisfy equation \ref{eq:ecc_orb_symmetry} and thus the EB model used to calculate the symmetrical LC point does not entirely correspond to the actual surface geometry. Therefore, conservative estimates of the photometric flux change are used to confirm sufficient similarity between the paired EB models. The estimates evaluate the effect of the difference between surface geometries and mutual irradiation.

The condition for the relative change in the surface geometry estimates the photometric flux change due to the difference in the component's point radius $\delta r_{n, m^{\prime}}$ between the orbital positions $n$ and $m^{\prime}$. As illustrated in Figure \ref{fig:orbit_symmetry}, the orbital position $m^{\prime}$ belongs to the orbital position couple containing the position $m$, which is itself a symmetrical counterpart of the position $n$ and thus sharing the same surface geometry. This estimate is then compared to the maximum allowed change in the flux $\delta F$ which is  $2\times10^{-4}$ in our case, the value adopted from the precision of the discretisation method used in ELISa:
\begin{align}
 \label{eq:geometry_constraint}
 \frac{2r_{eq}*\delta r_{n,m^{\prime}} + \delta r_{n,m^{\prime}}^2}{\sum r_{eq}^2} & \leq \delta F, \\ 
 \nonumber \delta r_{n,m^{\prime}} & = |r_{point, n} - r_{point, m^{\prime}}|, 
\end{align}+
where summation in the denominator goes through equivalent radii $r_{eq}$ of both components.

A change in the distance of the components between the orbital positions $n$ and $m^{\prime}$ affects the amount irradiation arriving from the binary companion to the component in question. Therefore, the following constraint is evaluated to limit the difference between fractions of the surface integrated irradiance $L_{irr}$ to the total luminosity of the system $L$:
\begin{equation}
 \label{eq:irradiation_constraint}
 \left|\left[ \frac{L_{irr}}{L}\right]_{m^{\prime}} - \left[\frac{L_{irr}}{L} \right]_n\right| \leq \delta F.
\end{equation}
Subsequently, the irradiation from the stellar companion $L_{irr}$ can be estimated as a fraction of the companion's luminosity captured by the irradiated star based on the components' distance $d$, equivalent radii $r_{eq}$, and effective temperatures of the components $T^{eff}$. Finally, the ratio $L_{irr}/L$ can be approximated for the `target' component $t$ while being irradiated by the `source' component $s$:
\begin{equation}
\label{eq:irradiation_condition}
 \frac{L_{irr}}{L} \approx
   \left(\frac{r_{eq, s}}{d}\right)^2 \left(1 + \left(\frac{r_{eq, s}}{r_{eq, t}}\right)^2 \left(\frac{T_s^{eff}}{T_t^{eff}}\right)^4 \right)^{-1}.
\end{equation}
If the condition in Equations \ref{eq:geometry_constraint} and \ref{eq:irradiation_constraint} are met, the orbital positions $n$ and $m^{\prime}$ are considered as similar.

\subsubsection{Similarity of adjacent orbital positions.}
\label{sec:third_approx}
Components of the binary stars with an eccentric orbit spend a large portion of the time near the apastra of their orbits. Therefore, observations equidistant in time tend to over-sample parts of the orbit with the maximum separation of the components, the slowest orbital motion, and the slowest change in surface geometry and mutual irradiation. Therefore, this approximation is meant to take advantage of this property, and it is based on the evaluation of the similarity of EB models on adjacent orbital positions. The similarity of EB models on two orbital positions is performed similarly to the process described in Section \ref{sec:second_approx}. This approximation proves extremely useful in cases with a high density of LC points and for EBs with an eccentricity close to zero. In such cases, surface geometry does not change rapidly or significantly during the orbital motion and thus recalculating the EB model for each orbital position is wasteful.

\subsubsection{Precision and speed of the approximations}
\label{sec:approx_pecision_vs_speed}
The approximation efficiency mentioned above significantly depends on LC points' density, eccentricity, and the filling factor of EB components. The interpolation approximation (Section \ref{sec:first_approx}) is slightly faster for the LCs with higher point densities. On the other hand, the symmetrical counterparts approximation described in Section \ref{sec:second_approx} performs well for EBs with orbital eccentricities above the value 0.6, where the interpolation approximation generally cannot be implemented due to a large angular separation of the orbital positions near the periastron. Additionally, the similar neighbours approximation can be stacked with either of the remaining methods. As a result, a combination of approximation methods provides a compounding effect on reducing computational time.

A benchmark based on a sample of 10\,000 randomly generated binary systems evaluated the performance of approximation methods. Orbital eccentricities and arguments of periastron were drawn from their full range, that is $e\in\langle 0, 1\rangle$, $\omega \in \langle 0, 360\rangle\degree$. The inclinations were selected from the interval $\langle 50, 90\rangle\degree$ and the mass ratio was restricted to the interval $(0, 1 \rangle$. Both surface potentials were selected from values below 50, with the lower boundary being the component's critical potential at periastron, where the component fills its Roche lobe. Discretisation factors of both components were set to 5 during the whole procedure. Subsequently, the LCs of such systems were generated using interpolation approximation and symmetrical counterparts approximation. LCs were calculated with a random number of surface points $N$ ranging from 100 to 900. LC points were generated equidistantly on the phase interval $\langle 0, 1\rangle$. Both methods were also supplemented with similar neighbours approximation as implemented in ELISa code. 

The resulting curves were then compared to the exactly calculated light curve of the same binary system. The left panel in Figure \ref{fig:ecc_lc_approx} illustrates the distributions of maximum deviations of approximated LCs from the observations calculated exactly for interpolation approximation and symmetrical counterparts approximation. Additionally, Figure \ref{fig:param_ecc_approx} provides a detailed view of the distribution of the residuals along the parameters $q$, $i$, $e$, $\omega$, and $N$ for both benchmarks. The results show that an overwhelming portion of the modelled LCs did not deviate from the exactly calculated reference LC by more than $2\times10^{-4}$ in a normalised photometric flux. On the other side, the right panel of Figure \ref{fig:ecc_lc_approx} points to the substantial improvement in the average computational speed if the approximations were utilised. Precision $2\times10^{-4}$ corresponds to the precision achieved by the currently used surface discretisation method determined in Section \ref{sec:lc_precision}. These results make this set of approximations, when applicable, a powerful tool for saving computational time in the case of EBs with eccentric orbits that usually require the largest amount of computational power.

\section{Visibility of surface elements during eclipse}
\label{sec:visibility}

\begin{figure}
        \centering
                \includegraphics[width=0.49\textwidth]{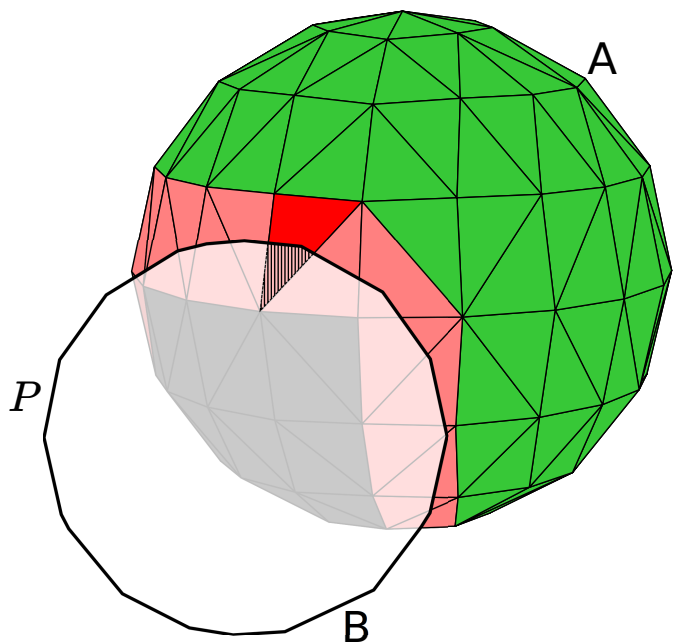}
                \caption{This diagram illustrates the process of determining the visibility of surface elements of star A, which is partially covered by star B during the eclipse. Green surface elements are fully visible to the observer, whereas grey elements are considered to be hidden. Pink (and red) are partially visible surface elements based on the algorithm described in Section \ref{sec:visibility}. The treatment of partially visible surface elements is demonstrated on the example of a surface element marked in red. A substantial increase in precision was achieved with the calculation of the overlapping area of the surface element marked in red. }
                \label{fig:eclipse}
\end{figure}

Modelling LCs of EBs requires an ability to evaluate the visibility of surface elements from the observer's viewpoint during the orbital motion. The determination of visibility for surface elements and establishing the component's horizons belong to one of the most challenging issues in the EB modelling, significantly affecting the observed flux during eclipses.
The following paragraph describes the solution to the visibility problem during an eclipse where the (partially) obscured EB component is referred to as star A and the component hiding star A is referred to as star B.

Initially, the EB model is rotated to align its new x-axis with the line of sight vector and surface points from both stars facing towards the observer are projected into $yz$ plane. Afterwards, polygon $P$ is found as a convex hull of projected points of star B, which later act as an outline of star B. As a result, the polygon $P$ represents the horizon of component B. Subsequently, it is determined whether the projected points of star A facing the observer lie inside the outline $P$. 
This procedure is performed by an efficient algorithm implemented in the matplotlib\footnote[1]{\url{https://matplotlib.org/3.1.3/api/path_api.html}} package \citep{hunter-2007}. If each point of the given surface element lies outside of the outline $P$, the surface element is fully visible to the observer. If each point of the surface element lies inside the outline $P$, the surface element is hidden behind star B, and its flux does not contribute to the LC integration. Finally, the surface elements containing points both inside and outside the outline $P$ are considered partially visible. They require additional attention, where the projected surface area visible to the observer is determined. The visible area can be determined indirectly by calculating the overlap between the projection of the surface element and the outline $P$. This process is illustrated in Figure \ref{fig:eclipse} where the overlap area of the partially eclipsed triangle is marked with stripes and the remaining visible part with a red colour.
Finally, the visible area corrected for its tilt with respect to the $yz$ plane can be used to calculate outgoing flux from the given surface element. As a result, the process described above significantly increases the numerical precision of the computed LC for the given level of the surface discretisation (i.e. the given number of surface elements). 

\section{Estimation of horizon}
\label{sec:horizon_est}
\begin{figure}
        \centering
                \includegraphics[width=0.49\textwidth]{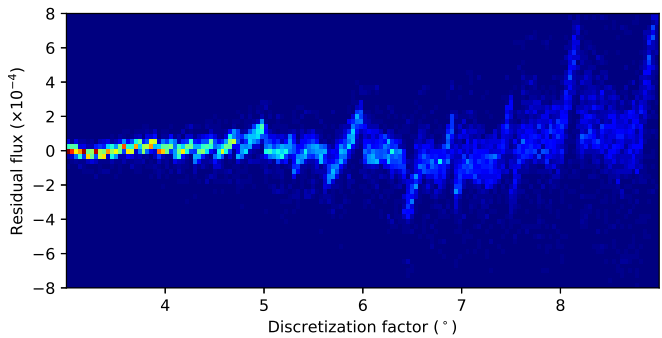}
                \caption{Dependence of emergent flux on the discretisation factor for 10\,000 randomly generated close eclipsing binaries.}
                \label{fig:solar_const}
\end{figure}

The surface discretisation process described in Section \ref{sec:surf_points} produces a set of surface elements (triangles) with their vertices located on an equipotential surface. However, the rest of the surface element is located underneath the desired surface, which underestimates the total area of the component and thus the total amount of outgoing flux. Moreover, the degree of the surface underestimation scales non-linearly with the discretisation factor. Therefore, the size of the binary model is expanded to offset the loss of the surface area caused by the discretisation using the following correction factor:
\begin{equation}
\label{eq:mesh_correction}
    C = \sqrt{\frac{E\alpha}{\sin{E\alpha}}}
,\end{equation}
where $\alpha$ is a discretisation factor and factor E accounts for the effect of non-equilateral triangles produced by discretisation methods described in Section \ref{sec:surf_points}. Parameter E changes primarily with the discretisation method and discretisation factor. The values of parameter E were calibrated on a statistically significant sample of randomly generated binary systems. The calibration process demonstrated that parameter E does not change significantly with surface geometry. The main reason for this behaviour is the deterministic nature of the algorithms for generating surface mesh. A sample of 10\,000 detached and overcontact EBs with randomly generated parameters and observational phases was used to investigate a relationship between the relative flux and used discretisation factor. Obtained photometric fluxes were then normalised to the reference values calculated for the discretisation factor 1. The corrected fluxes obtained from our test sample are displayed in Figure \ref{fig:solar_const}. The results demonstrate the independence of the photometric fluxes on the used discretisation factor. 

\begin{figure}
        \centering
                \includegraphics[width=0.49\textwidth]{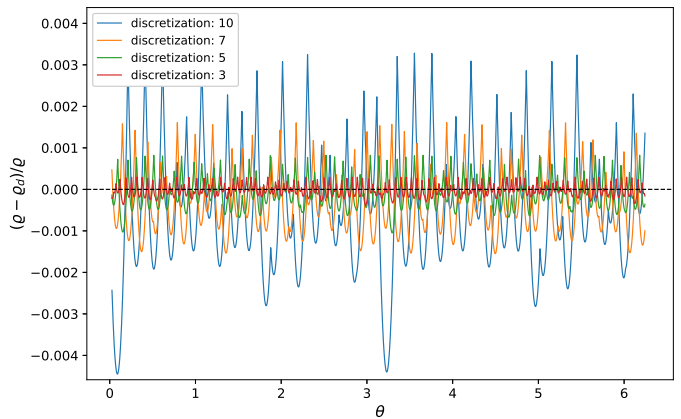}
                \caption{Estimation of the stellar horizon using discretisation factors 10, 7, 5, and 3. A lower discretisation factor leads to denser coverage of surface points and, thus, a more precise estimation of the horizon. The horizon was obtained at the photometric phase 0.1 and inclination 60\degree.}
                \label{fig:horizon_vs_disckretization}
\end{figure}

Due to the nature of the surface discretisation, the precision of the component horizon significantly depends on the discretisation factor. The surface defined using a smaller discretisation factor copies the true horizon much more closely, as demonstrated in Figure \ref{fig:horizon_vs_disckretization}. Based on the results from this section, discretisation factor 5 was picked as an acceptable compromise between precision and speed.

We focused the subsequent analysis on the horizon estimation by investigating the effects of seams produced by the discretisation method used by ELISa. This issue was investigated by plotting the residuals of the synthetic horizon against the true horizon for several different phases and inclinations along the seams using discretisation factor 5 (see Figure \ref{fig:horizon}). Subsequently, a standard deviation of the residuals was calculated and listed in Table \ref{table:horizon_vs_inclination}. Even though the shapes of residuals show the expected symmetry in residuals for certain viewing angles, the consistent values of the standard deviations across Table \ref{table:horizon_vs_inclination} suggest that the surface discretisation method does not significantly affect the precision of synthetic observations.

\begin{table}
\caption{Precision in the estimation of the horizon by the discretisation method used by ELISa. Mean values and standard deviations of the relative precision of the horizon $\delta \rho = (\varrho - \varrho_d) / \varrho$ were evaluated along the equator and meridian. Values of means and standard deviations are in order of $10^{-4}$.}
\label{table:horizon_vs_inclination}
    \centering
        \begin{tabular}{c c c c c}
        \hline\hline
        Inclination ($^\circ$) & 80 & 60 & 30 & 0  \\    
        \hline
        $\overline{\delta \rho}$ & -0.9  & -1.2 & -1.2 & 0.7 \\
        $\sigma(\delta \rho)$ & 3.3  & 3.2 & 3.6 & 3.6 \\
        \hline
        Phase & 0.0 & 0.15 & 0.25 & 0.45  \\    
        \hline
        $\overline{\delta \rho}$ & -1.1  & -0.6 & 1.7 & -1.2\\
        $\sigma(\delta \rho)$ & 4.6  & 3.5 & 3.8 & 3.6\\
        \hline
        \end{tabular}
\end{table}

\section{Inverse problem}
\label{sec:inverse problem}
One of the ultimate goals of  ELISa is to perform as a solver for an inverse problem where system parameters and their confidence intervals are inferred from the observations. To achieve this goal, observed LC or RV data used during the fit are processed in ELISa in the form of the dedicated data set module, which contains $n$ observations $y^{obs}$ with corresponding times $t$ and observational errors $\sigma$. On the other hand, the optimisation method generates synthetic observations $y^{mod}(\boldsymbol{x})$, which are defined using a combination of EB parameters $\boldsymbol{x}$. The goodness of the model $\boldsymbol{x}$ is then evaluated by a suitable cost function.

A user can supply a combination of EB parameters $\boldsymbol{x}$ either in a 'standard' format containing masses of components or in a 'community' format instead by providing the system's mass ratio and the semi-major axis of the system. Each parameter of the EB system can be set as a variable, fixed, or constrained parameter. Variable parameters change during the fit, and the optimizer method adjusts them to find optimal solutions to the inverse problem. The user needs to set fixed variables at the start, and they subsequently remain constant during the optimisation process. Finally, constrained parameters change during optimisation by evaluating the user's constraint statement containing any number of variable parameters and allowed operators.

A solution to the inverse problem can be achieved by using various fitting methods. The desired process should provide the solution in the least amount of time (i.e. require the least amount of model evaluations) and robustly estimate the uncertainties of the found parameters. In our experience, the inverse problem can be split into two stages, where different methods are more suitable. The first stage consists of finding an acceptable solution that minimises the cost function. For this reason, ELISa utilises a commonly used non-linear large-scale LSTRR algorithm, which enables a quick search for local minima inside a predefined search box using a gradient method. Subsequently, once a local minimum is found, a stochastic method such as MCMC can be used to sample the parameter space in the vicinity of the solution and to provide robust estimations of EB parameters. Additionally, the quality of the obtained solution can be examined by the coefficient of determination $R^2$. Details on the implementation of each optimisation method in ELISa are discussed in the following sections.

\subsection{LSTRR}
\label{sec:lstsqr}
ELISa uses the LSTRR\footnote[1]{\url{https://docs.scipy.org/doc/scipy/reference/generated/scipy.optimize.least_squares.html}} algorithm implemented in SciPy \citep{2020SciPy-NMeth}. The LSTRR algorithm utilises a gradient approach to search for a local minimum of the weighted sum of squares of the residuals $\xi^2$:
\begin{equation}
\label{eq:xi2}
\xi^2 = \sum_{i=1}^{n} \frac{[y^{obs}_i - y^{mod}_i(\boldsymbol{x})]^2}{\sigma_i^2}.
\end{equation}
Unfortunately, the success rate of the method significantly depends on the initial parameters. Thus initial input from the user is required, thus, making utilisation of the LSTRR method alone in any automatised application very difficult.

Due to the complex nature of the LC evaluator in ELISa and LC evaluators in general, numerous degenerate states of the LC evaluator need to be addressed by penalising such models by setting a $\xi^2$ function to practical infinity. Such a penalisation is performed in cases involving nonphysical parameters, system parameters outside the supported range (no atmospheric or limb darkening model available), or undesired morphology of the system.

\subsection{MCMC}
\label{sec:mcmc}
ELISa also uses a pure Python implementation of the affine-invariant MCMC ensemble sampler \citep{foreman-mackey} available in the emcee\footnote[1]{\url{https://emcee.readthedocs.io/en/stable/}} package. In this case, the likelihood function was defined using an assumption that the observational data are drawn from a Gaussian distribution around the actual values. In our case, we defined the likelihood function describing logarithm of the probability of the observations $y^{obs}$ to be from the Gaussian distribution described by the mean $y^{mod}$ and the variance $\sigma$ as:
\begin{equation}
\label{eq:lhood}
\ln{p(y^{obs}|y^{mod},\sigma)} = -\frac{1}{2}\sum_i^n \left\{ \frac{[y^{obs}_i - y^{mod}_i(\boldsymbol{x})]^2}{\sigma_i^2 + s_i^2} + \ln{[2\pi (\sigma_i^2 + s_i^2)]} \right\}, 
\end{equation}
where $s_i$ is an error underestimation parameter that accounts for the fact that, in general, the synthetic model does not explain 100\% of the variability present in the observations. Parameter $s_i$ can be expressed as follows:
\begin{equation}
\label{eq:error_underestimation}
s_i = \log{(f)} y^{mod}_i(\boldsymbol{x}),
\end{equation}
where $\ln{f}$ is a marginalisation parameter fitted alongside the rest of the variables.
As for version 0.5, ELISa supports the sampling from uniform or normal prior distribution, which can be used to utilise the prior knowledge about the fitted parameter. Parameters with uniform priors require one to specify the upper and lower boundary of the sampling. On the other hand, the parameters with normal prior distribution are drawn around the most probable value with the expected standard deviation. Additionally, the normal prior distribution can be bounded with the lower and upper boundary, similarly to the uniform distribution to prevent the sampler from reaching the undesired areas of the parameter space.

As in the case of the LSTRR method in Section \ref{sec:lstsqr}, degenerate states of the sampler have to be treated. Invalid sampler states are treated by assigning very low values of the likelihood function to penalise a combination of parameters, leading to non-physical systems or systems outside the support of pre-calculated tables.

The resulting chains are then processed, and the most probable values, along with their confidence intervals, are calculated. The resulting constrained parameters are calculated using the most probable values of variable parameters obtained from the MCMC sampler. ELISa also provides valuable tools for managing the output chain, such as discarding the samples from the initial thermalisation stage or filtering the chain to the desired interval according to any variable parameter to deal with multiple peak solutions. Users can also use the resulting chain to propagate errors of the fitted parameters to the other derived parameters (e.g. inferring uncertainty of the component radius based on the surface potential and the system's mass ratio).

\subsection{General guidelines for solving the inverse problem}
To speed up the process and to ensure the quality of the estimated parameters, a few guidelines are provided: 

{\it Reduce the number of fitted parameters.} To increase the probability that the optimizer will find the optimal solution and reduce computational time, the user should keep the number of the fitted parameter at a minimum. Parameters that do not significantly affect the shape of the synthetic curve (e.g. gravity darkening factor, albedo, and metallicity) should be kept fixed at some reasonable value during the initial runs. Subsequently, when a rough solution is found, the rest of the parameters can be set as a variable to refine the solution. Parameters, such as ephemeris that can be obtained from less computationally demanding methods, should be calculated a priori and kept fixed during the fit.

{\it Combine methods.} The LSTRR and MCMC methods have their advantages and disadvantages, as is discussed in Sections \ref{sec:lstsqr} and \ref{sec:mcmc}. The LSTRR algorithm is far more efficient in minimising the cost function. Therefore, the LSTRR should be used to search for a local minimum around the initial values. Subsequently, the parameter space near the found solution can be sampled using the MCMC method to estimate confidence intervals of fitted parameters. Additionally, the MCMC algorithm performs badly as a minimizer. It requires significantly longer computational time to generate a chain usable for the analysis of posteriors if used as a stand-alone fitting method.

\section{Multiprocessing}
ELISa supports a multiprocessing approach to the EB modelling on two levels. The first multiprocessing approach is implemented during the LC integration, where the requested LC points are divided into batches that are then evaluated as separate processes. The second approach is utilised either during the MCMC method or used anywhere where a large number of independent synthetic observations are produced. In such a case, multiple LCs are calculated in parallel. The second parallelisation approach should be used where possible since it brings the least amount of overhead and thus is more efficient. Users should always avoid the usage of both parallelisation techniques simultaneously.

\section{Runtime and precision}
\label{sec:precision}

One of the main goals of ELISa is to provide the EB modelling capability with balanced speed and precision. Sufficient precision is necessary to be able to process the latest high precision space-based observations. On the other hand, using such a modelling tool in optimisation methods and other automatised applications requires an appropriate runtime speed so a large amount of synthetic data can be generated. 

\subsection{Precision of the light curves}
\label{sec:lc_precision}
\begin{figure}
        \centering
                \includegraphics[width=0.49\textwidth]{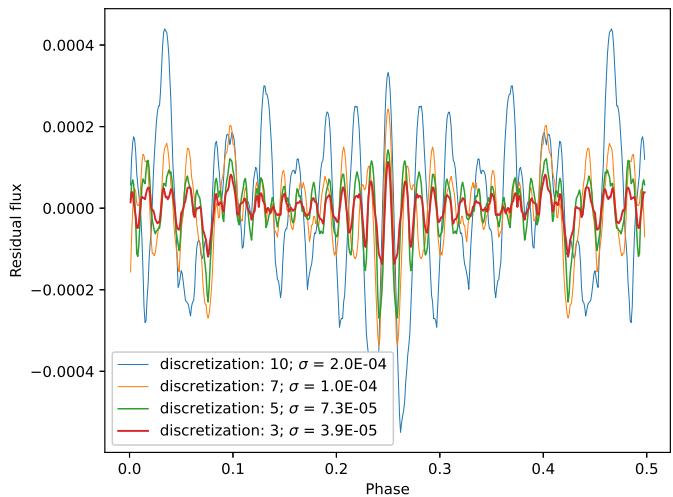}
                \caption{Numerical noise observed in the synthetic observations due to the surface discretisation. Residuals of the synthetic observations for discretisation factors 10, 7, 5, and 3 are displayed in different colours along with their standard deviations $\sigma$. The amplitude of residuals for discretisation factor 5 is roughly $2\times10^{-4}.$ This value should be regarded as the precision of the numerical model for this level of surface discretisation. Due to the symmetry of the surface discretisation, only one half of the phase curve is displayed for the sake of clarity.}
                \label{fig:lc_precision}
\end{figure}

We performed several benchmarks to evaluate the precision and the runtime speed of this package. Initially, the amount of noise in the synthetic data was determined using an LC of a very wide binary system with an inclination of 90\degree \ and components sufficiently compact to prevent any ellipsoidal variation. Such a benchmark was supposed to examine the noise caused by the surface discretisation and numerical operations. Such an EB model resulted in the LC that was practically flat and suitable for studying the noise levels in a bolometric filter. LCs for different values of the surface discretisation parameter are displayed in Figure \ref{fig:lc_precision}. The residuals show phase-correlated variations caused by the surface discretisation. However, the amplitude of residuals for discretisation factor 5 reached $2\times10^{-4}$, which we consider to be an acceptable compromise between computational speed and precision. 

\subsection{Performance}
\label{sec:performance}
We performed another set of benchmarks to evaluate the performance and precision of ELISa by evaluating the LCs of EBs with circular and eccentric orbits. Because computational speed highly depends on the used hardware and evaluating the precision of the synthetic LC is problematic without any reference, we compared the results of the benchmark models obtained by ELISa to the results obtained by PHOEBE v2.2, which is considered as the standard tool for EB modelling. 

\begin{figure}
        \centering
                \includegraphics[width=0.49\textwidth]{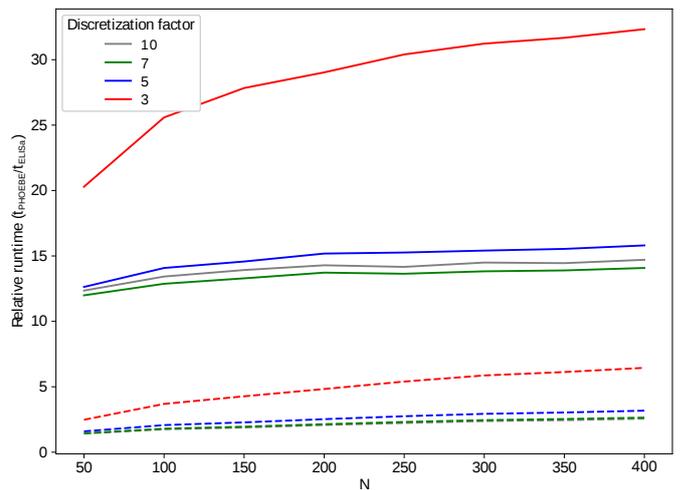}
                \caption{Relative runtime of ELISa while evaluating the LCs in three different passbands for EB models with circular (solid lines) and eccentric orbit (dashed lines) with the eccentricity of $e=0.2$ (see model parameters in Table \ref{table:models}). The LCs were obtained for different values of the surface discretisation factor and the number of points $N$ on the LC. We then compared the results obtained by ELISa with PHOEBE v2.2 performance on the same benchmark models.}
                \label{fig:elisa_vs_phb}
\end{figure}

We evaluated the EB models (see Table \ref{table:models}) in both modelling software packages. Additionally, the comparability of the results was ensured by setting the same number of surface elements for both runs. Figure \ref{fig:elisa_vs_phb} illustrates the effect of using surface symmetries and approximations during the LC integration described in sections \ref{sec:surf_points} and \ref{sec:lc_integration} where the single-core performance of ELISa significantly increased.

However, it has to be noted that such a performance boost is gained when surface symmetries or approximations can be used. In the case of circular orbits, a significant performance boost is gained if the same EB model can be used to calculate fluxes at all phases. Similarly, in cases of EBs with eccentric orbits, ELISa does not achieve the level of performance as displayed in Figure \ref{fig:elisa_vs_phb} if the system contains spots or any surface irregularity.  Such a feature would have to be repositioned in the co-rotating frame of reference during the LC integration. The presence of the spot itself also slightly affects the performance since the package cannot fully utilise the surface symmetry. Additionally, saved computational time significantly depends on overall changes in geometries of the components during the orbital motion. Therefore, the best performance boost is gained for low eccentricity systems or systems with compact components, well inside their respective Roche lobes. 

\subsection{Light curve comparison}
\label{sec:LC_comparison}

Finally, we compared the LCs calculated by both codes for each benchmark model. Both codes used the atmosphere tables from \citet{castelli01}, and logarithmic law was used for the calculation of the limb darkening. Limb darkening coefficients used by PHOEBE were set manually to average values used by ELISa to ensure comparability of the results.

Comparison of the obtained result (displayed in Figure \ref{fig:elisa_vs_phoebelc}) shows that both circular and eccentric benchmark models show discrepancies with the amplitude of $\approx 10^{-3}$, above the relative numerical precision stated in Section \ref{sec:lc_precision}. We identified the main source of the phase correlated residuals to be the treatment of bolometric and passband-related limb darkening coefficients set in PHOEBE as a constant across the surface and during the orbital motion. The secondary variations are more stochastic, and their amplitude peaks at $2\times10^{-4}$. These variations are mainly caused by the surface discretisation used by ELISa, along with approximations taken while integrating the LC of the eccentric benchmark model.

\section{Conclusions}
ELISa is a new software package dedicated to modelling EBs written in pure Python utilising modern techniques in computational astrophysics. Modelling EB systems applies up to date methods of modelling stellar surfaces based on the triangulation of surface points which improves the precision of the model compared to the older WD code. In addition, the integration of the synthetic observations utilises surface and orbital symmetries to significantly reduce the computational time necessary to build the EB model and calculate the observed quantities (sections \ref{sec:surf_points} and \ref{sec:lc_integration}). This package can also model temperature spots using dedicated surface discretisation capable of spot stacking and modelling small spots compared to surface discretisation. Additionally, support for the modelling of non-radial pulsations will be added soon. ELISa also provides the built-in capability to solve an inverse problem using the LSTRR algorithm and MCMC as discussed in Section \ref{sec:inverse problem}. The implementation of the inverse task module enables one to fix or constrain an arbitrary EB parameter. Finally, in Section \ref{sec:precision}, we evaluated the precision and computational speed of this package by comparing the results and its performance with the standard EB modelling package PHOEBE. We demonstrate that ELISa balances precision with computational speed, making it an excellent tool for applications requiring the evaluation of a large amount of EB models such as solving an inverse problem or generating large datasets for applications such as machine learning.

Users have to also be aware of the limitations of the ELISa package. The most obvious one is the relative precision of the synthetic observations currently limited to $2\times10^{-4}$. This precision level is by order of magnitude worse than the accuracy of the available space-based observations by Kepler \citep{Borucki16}. The main culprit here is the discretisation method, where the effect of seams on the equator and $\varphi = 0, \pi/2$ meridians was not fully rectified. The presence of the seams is the direct consequence of the utilisation of surface symmetry. However, the authors believe that this package's current level of precision is absolutely adequate to derive sound estimates of the EB parameters and their confidence intervals. The ELISa package (version 0.5) also lacks the capability to model phenomena such as the apsidal motion or the orbital and rotational Doppler beaming effect \citep{Hills74, Maxted00}. Last but not least, a shortcoming is currently the lack of the graphical user interface (GUI) that we tried to remedy with the design of easy to use scripting capabilities.

In conclusion, ELISa is primarily a fast and easy to use EB modelling package with significant scripting capabilities that require only rudimentary programming skills to operate. The capabilities described in this paper make ELISa the ideal tool for simple, robust, and time-efficient inference of EB parameters from observational data without the users needing to develop their own or adapt already existing optimisation algorithms. On the other hand, ELISa can also be readily used as a test bed in EB research, especially for more complex applications requiring the evaluation of a significant amount of synthetic observations.

\section*{Acknowledgements}
This research was supported by the Slovak Research and Development Agency under contracts No. APVV-15-0458 and APVV-20-0148. The research of M.F. was supported by the internal grant No. VVGS-PF-2019-1392 of the Faculty of Science, P. J. {\v S}af{\'a}rik University in Ko{\v s}ice.

\bibliographystyle{aa}
\bibliography{ref}

\begin{appendix}
\onecolumn

\section{Performance of eccentric orbit approximations}
\label{app:ecc_approx}
\begin{figure}[h!]
\begin{center}
 \includegraphics[width=0.8\linewidth]{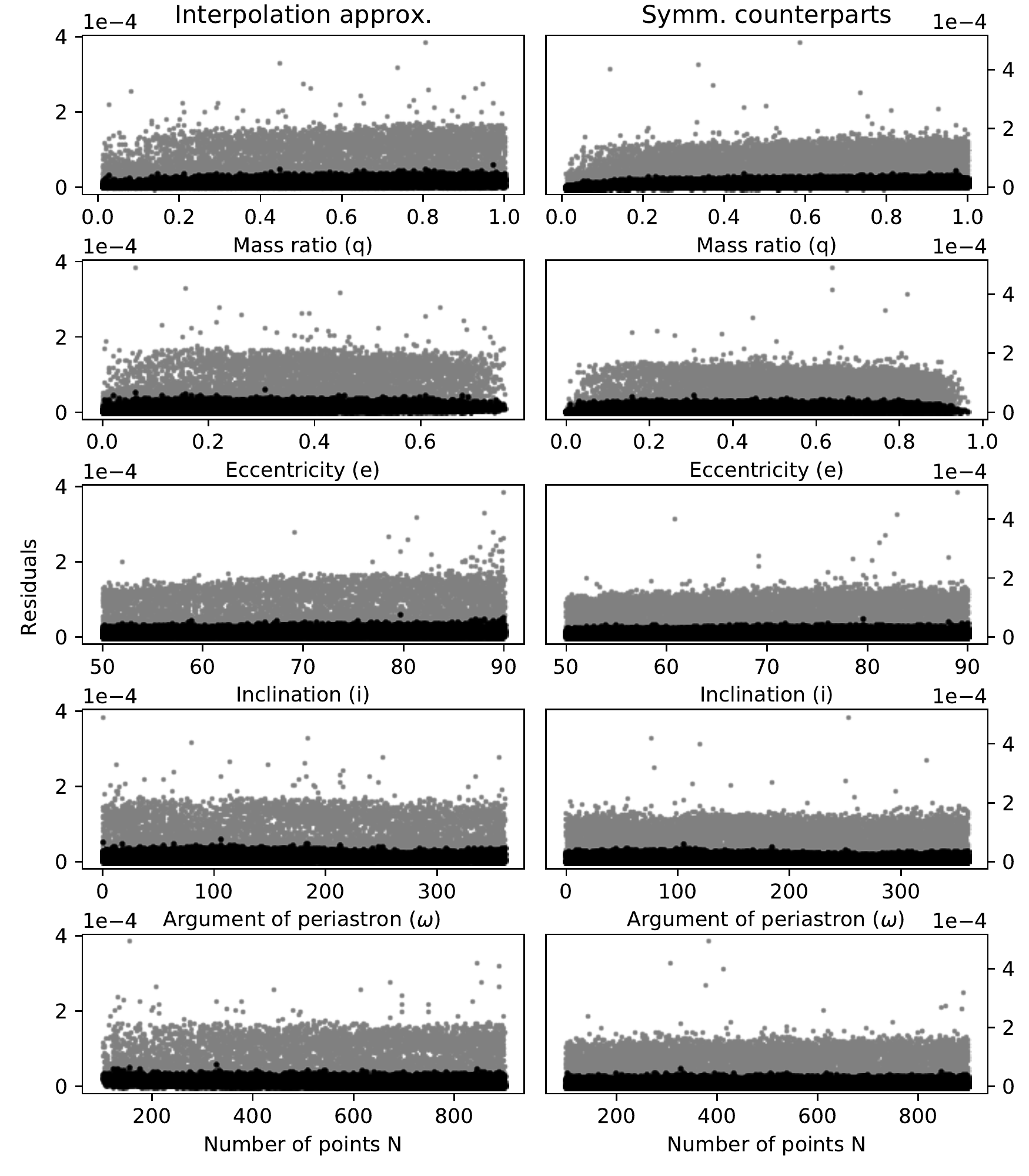}
\end{center}
 \caption{Maximum deviations (grey points) and average deviations (black points) of the approximated light curves from the exactly calculated control sample. The panels illustrate distributions of the residuals within the parameter space from which we generated the 10\,000 random binary systems. The left column displays the result of the combination of interpolation (Section \ref{sec:first_approx}) and the similar neighbours approximation (Section \ref{sec:third_approx}), whereas the right column shows the results for the symmetrical counterpart (Section \ref{sec:second_approx}) and similar neighbour approximation.}
\label{fig:param_ecc_approx}
\end{figure}

\newpage
\section{Light curve comparison}
\label{app:benchmark}
\begin{table}[h!]
    \caption{Parameters of EB models used for benchmark tests in Section \ref{sec:precision}.}
    \label{table:models}
    \centering
        \begin{tabular}{c c c c c c c c c c c c c c c c c}
        \hline\hline
        model & $e$ & $i[^{\circ}]$ & $\omega[^{\circ}]$ & $P[d]$ & component  & $M[M_{\odot}]$ & $R_{eq}[R_{\odot}]$ & $\Omega$ & $T_{eff}[K]$ & $F$ & $A$ & $\beta$   \\ 
        \hline
        \multirow{2}{*}{circular} & \multirow{2}{*}{0.0} & \multirow{2}{*}{80.0} & \multirow{2}{*}{0.0} & \multirow{2}{*}{1.2} &
        
        \multirow{1}{*}{primary} & \multicolumn{1}{c}{2.0} & \multicolumn{1}{c}{2.380} & \multicolumn{1}{c}{3.8} & \multicolumn{1}{c}{7000} & \multicolumn{1}{c}{1.389} & \multicolumn{1}{c}{1.0} & \multicolumn{1}{c}{1.0} \\
        
        & & & & & \multirow{1}{*}{secondary} & \multicolumn{1}{c}{1.4} & \multicolumn{1}{c}{1.147} & \multicolumn{1}{c}{4.8} & \multicolumn{1}{c}{6300} & \multicolumn{1}{c}{1.5} & \multicolumn{1}{c}{0.6} & \multicolumn{1}{c}{0.32} \\
        \hline
        \multirow{2}{*}{eccentric} & \multirow{2}{*}{0.2} & \multirow{2}{*}{80.0} & \multirow{2}{*}{170.0} & \multirow{2}{*}{1.9} &
        
        \multirow{1}{*}{primary} & \multicolumn{1}{c}{2.0} & \multicolumn{1}{c}{1.811} & \multicolumn{1}{c}{6.0} & \multicolumn{1}{c}{7000} & \multicolumn{1}{c}{1.5} & \multicolumn{1}{c}{1.0} & \multicolumn{1}{c}{1.0} \\
        
        & & & & & \multirow{1}{*}{secondary} & \multicolumn{1}{c}{1.3} & \multicolumn{1}{c}{1.073} & \multicolumn{1}{c}{7.0} & \multicolumn{1}{c}{6000} & \multicolumn{1}{c}{1.0} & \multicolumn{1}{c}{0.6} & \multicolumn{1}{c}{0.32} \\
        \hline
        \end{tabular}
\end{table}

\begin{figure}[h!]
        \centering
            \includegraphics[width=0.49\textwidth]{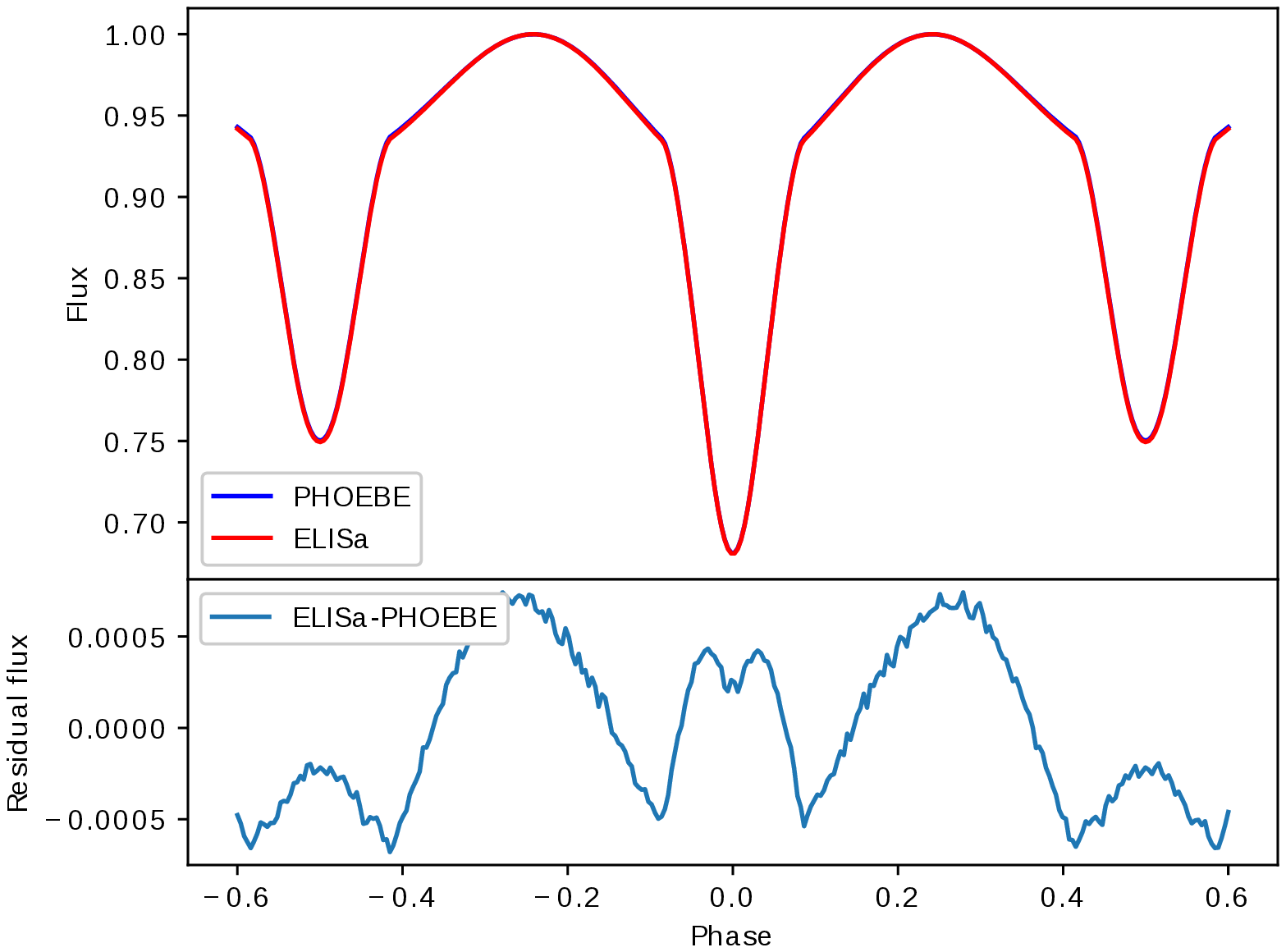}
            \includegraphics[width=0.49\textwidth]{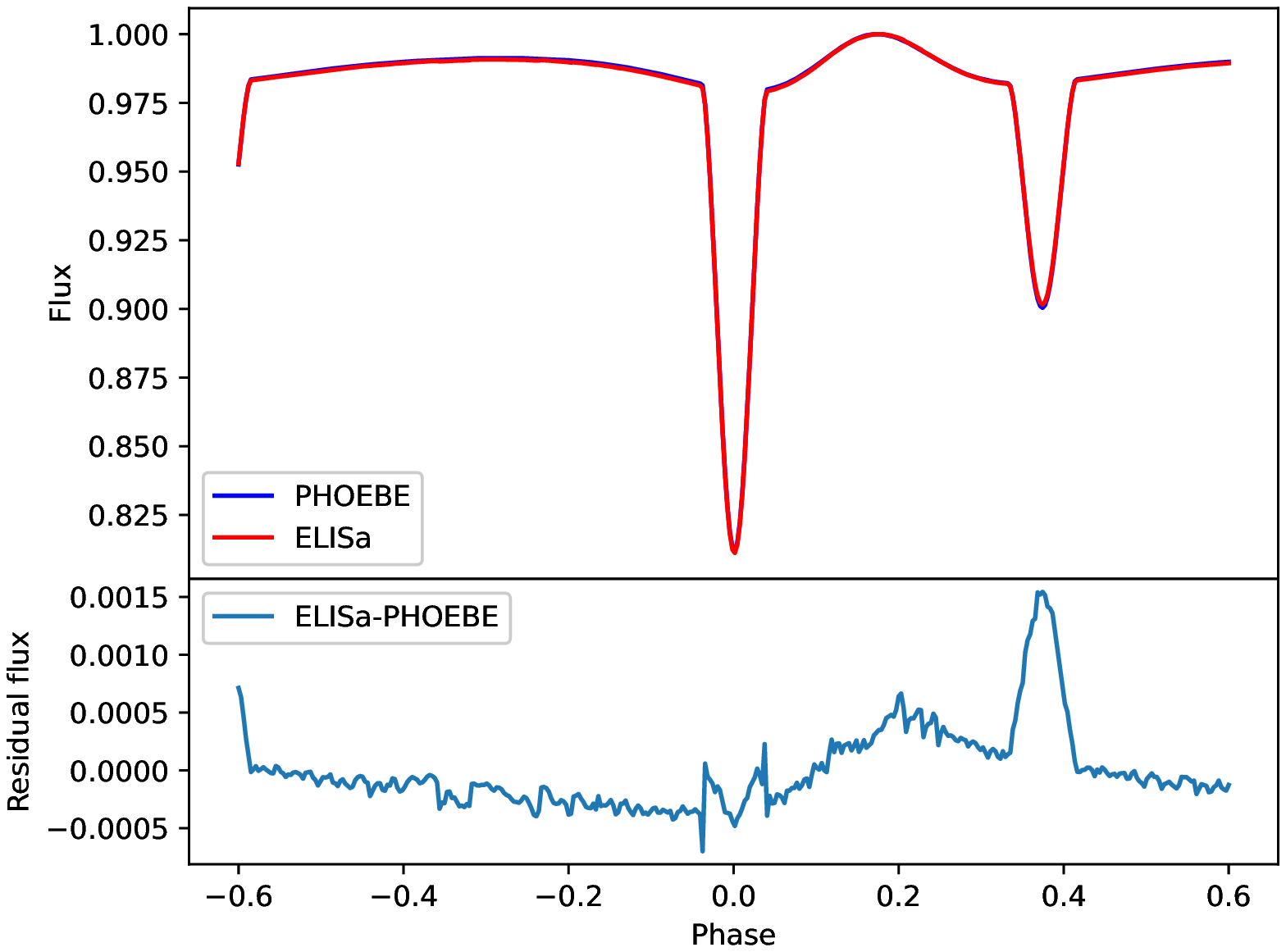}
                \caption{Light curves of two EB models calculated by ELISa and PHOEBE using the same number of approximately 3380 surface elements on the primary component. The figure on the left shows the light curve of the EB model with a circular orbit and the LC for the eccentric EB model on the right side. The shape of the residuals did not change significantly with a different discretisation factor.}
                \label{fig:elisa_vs_phoebelc}
\end{figure}

\section{Horizon determination}
\label{appendix:horizon}
\begin{figure}[h!]
        \centering
            \includegraphics[width=0.49\textwidth]{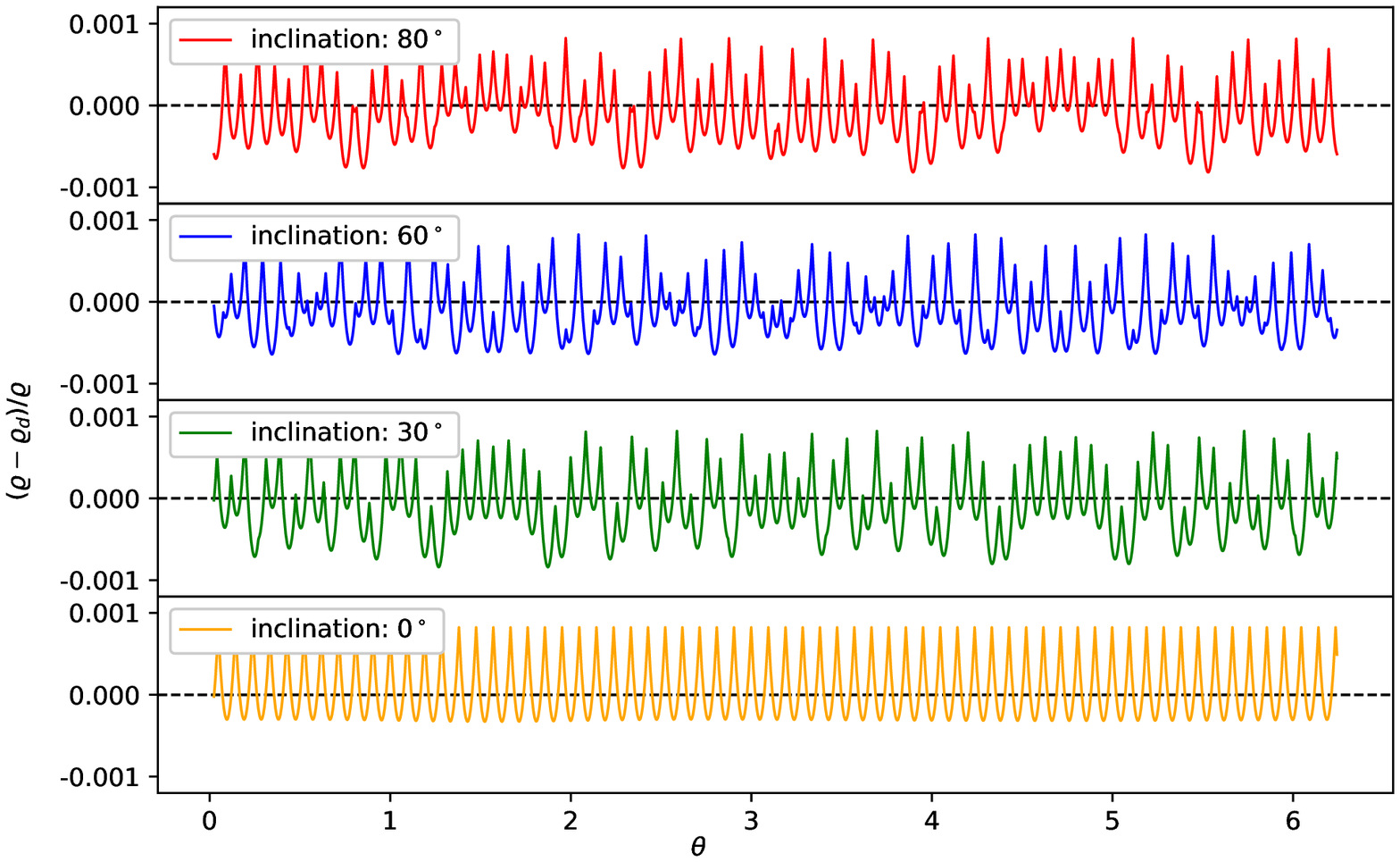}
            \includegraphics[width=0.49\textwidth]{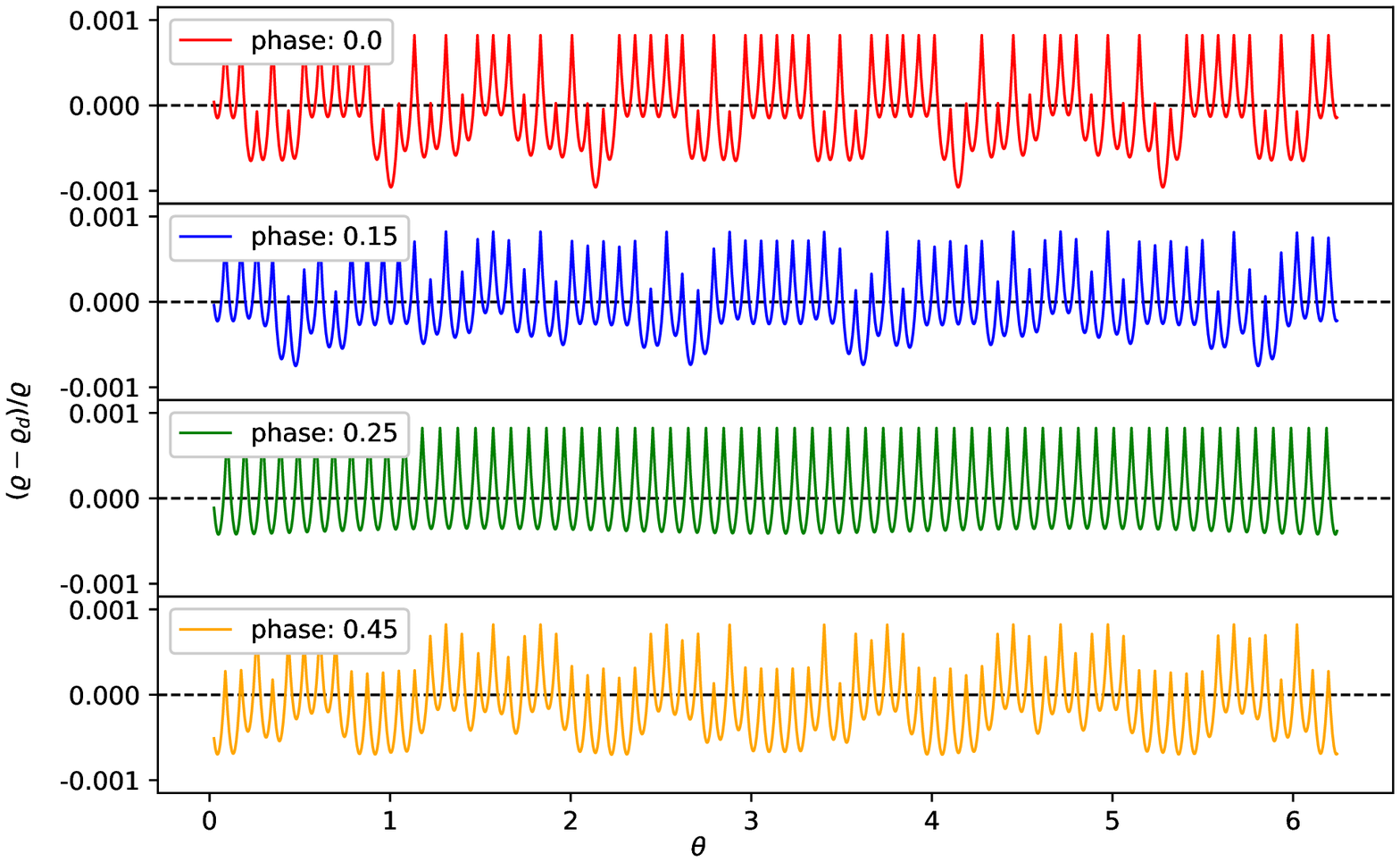}
            
                \caption{Estimations of stellar horizon along the discretisation seams. The figure on the left shows horizon estimations for photometric phase 0 and inclinations 80\degree, 60\degree, 30\degree, and 0\degree. The figure on the right focusses on the horizontal seam with horizon estimations for inclination 90\degree \ and photometric phases 0, 12, 25, and 45. The estimation of the horizon corresponding to the inclination 0\degree \ and photometric phase 0.25 shows symmetrical residuals due to the seam located directly on the horizon.}
                \label{fig:horizon}
\end{figure}

\end{appendix}
\end{document}